\documentclass[twoside,british,a4paper,twocolumn]{article}

\usepackage{amsmath}
\usepackage{amsfonts}
\usepackage{amssymb,upref}
\usepackage{tgheros} 
\usepackage{tgtermes} 
\usepackage{anyfontsize}
\usepackage{xcolor}
\usepackage{graphicx}
\usepackage{enumitem}
\usepackage{subfig}
\usepackage{siunitx}

\usepackage{balance}

\usepackage[utf8]{inputenc}
\usepackage[T1]{fontenc}

\usepackage{isodate}
\cleanlookdateon

\definecolor{refc}{rgb}{0.17, 0.402, 0.631}
\definecolor{msc}{rgb}{0., 0., 0.}
\usepackage{lettrine}
\pdfmapfile{=montserrat.map} 

\usepackage{lipsum}

\usepackage{newtxmath}

\usepackage{geometry}
\usepackage{fancyhdr}

\fancypagestyle{firstpagestyle}
\usepackage{titlesec}
\titleformat{\section}[block]{\large\bfseries\upshape\sffamily\boldmath}{}{0.em}{}
\titlespacing*{\section}{0pt}{0.8em plus 0ex minus 0ex}{0em plus 0.ex}
\titleformat{\subsection}[block]{\bfseries\upshape\sffamily\boldmath}{}{0.em}{}
\titlespacing*{\subsection}{0pt}{0.8em plus 0ex minus 0ex}{0em plus 0.ex}

\usepackage{titling}
\pretitle{
\begin{minipage}{\textwidth}\noindent \raggedright\huge\bfseries\upshape\sffamily\boldmath
\color{msc}
 }
\posttitle{ 
\end{minipage}\vskip 0.7em}

\preauthor{\noindent \begin{minipage}{\textwidth} \large\mdseries\sffamily}
\postauthor{
  \end{minipage} 
  \vskip 0.4em
  {
   \sffamily
   \color{black!80}
   \noindent \small
   \address
   \par 
   \authoremail
   }
   \par \vskip 0.4em
  }
\predate{ \noindent \hspace{-0.4em} \sffamily\small}
\postdate{\vskip 0.0em}

\usepackage{caption}
\DeclareCaptionFont{cfs}{\fontsize{8.5}{10.25}\selectfont}
\DeclareCaptionLabelSeparator{vline}{\;|\;}
\newcommand{\figurecaption}[2]{\caption[#1]{\textbf{#1.} #2}}

\usepackage{tcolorbox}
\definecolor{abstractboxcolor}{cmyk}{0.1,0,0,0}
\newtcolorbox{abstractbox}{
  arc=0pt,
  boxrule=0pt,
  colback=abstractboxcolor,
  boxsep=0.5em,
  left=0pt, right=0pt, bottom=0pt, top=0pt,
  width=\columnwidth
}
\makeatletter
 \def\@textbottom{\vskip \z@ \@plus 1pt}
 \let\@texttop\relax
\makeatother
\renewenvironment{abstract}{
   \noindent
   \begin{minipage}{\textwidth}
   \upshape\sffamily \bfseries
   \fontsize{9}{11.5}\selectfont
  }{
   \end{minipage} 
   \vskip 2.0em
  }

\usepackage[numbers,sort&compress]{natbib}

\makeatletter \def\NAT@def@citea{\def\@citea{\NAT@separator\,}} \makeatother 
\usepackage{etoolbox}
\apptocmd{\sloppy}{\hbadness 10000\relax}{}{}

\newcommand{\citers}[1]{Refs.~\citealp{#1}}

\newcommand{\sref}[2]{\ref{#1}{\color{refc} #2}}

\newcommand{\refedfig}[1]{Extended Data Fig.~\ref{#1}}

\usepackage{multibib}
\newcites{supp}{References}

\usepackage[%
  bookmarks=true,
  colorlinks,
  linkcolor=refc,
  urlcolor=refc,
  citecolor=refc,
  plainpages=false,
  pdfpagelabels,
  final,
  breaklinks=true
]{hyperref}

\usepackage{physics}
\newcommand{\ten}[1]{\boldsymbol{\mathcal{#1}}}
\newcommand{\bs}{\boldsymbol}
\newcommand{\bt}{\textbf}

\newcommand{\bse}{\begin{subequations}}
\newcommand{\ese}{\end{subequations}}

\newcommand{\Ein}{e^{(\text{in})}}
\newcommand{\Eout}{e^{(\text{out})}}
\newcommand{\bEin}{\bt e^{(\text{in})}}
\newcommand{\bEout}{\bt e^{(\text{out})}}

\graphicspath{{Figures/}}

\usepackage{textcomp}
\usepackage{dsfont}

\title{Crashing with disorder: 
Reaching the precision limit 
with tensor-based wavefront shaping
}

\newcommand\shorttitle{Crashing with disorder}

\author{Rodrigo Guti\'{e}rrez-Cuevas,\textsuperscript{1$\,*$} 
Dorian Bouchet,\textsuperscript{2} Julien de Rosny,\textsuperscript{1}  and 
S\'{e}bastien M. Popoff\textsuperscript{1}}
\newcommand\shortauthor{R. Guti\'{e}rrez-Cuevas, D. Bouchet, 
  J. de Rosny, and S.M.Popoff}

\newcommand\address{\textsuperscript{1}Institut Langevin, ESPCI Paris, Université PSL, CNRS, 
75005 Paris, France\\
\textsuperscript{2}Universit\'e Grenoble Alpes, CNRS, LIPhy, Grenoble, France}
\newcommand\authoremail{$^*$rodrigo.gutierrez-cuevas@espci.fr}

\begin{document}

\twocolumn[
\begin{@twocolumnfalse}

\maketitle
\thispagestyle{firstpagestyle}

\begin{abstract} 
Perturbations in complex media, due to their own dynamical 
evolution or to external effects, are often seen as detrimental. 
Therefore, a common strategy, especially for telecommunication and imaging applications, 
is to limit the sensitivity to those perturbations in order to avoid them.
Here, we instead consider crashing straight into them in order to maximize 
the interaction between light and the perturbations and thus 
produce the largest change in output intensity. 
Our work hinges on the innovative use of tensor-based techniques, 
presently at the forefront of machine learning explorations, to study intensity-based 
measurements where its quadratic relationship to the field prevents the use of standard 
matrix methods. With this tensor-based framework, 
we are able to identify 
the \emph{optimal crashing channel} which maximizes the change in its output 
intensity distribution and the Fisher information encoded in it about a given 
perturbation. 
We further demonstrate experimentally its superiority for robust and 
precise sensing applications.
Additionally, we derive the appropriate strategy to 
reach the precision limit for intensity-based measurements 
leading to an increase in Fisher information 
by more than four orders of magnitude with respect to the mean for
random wavefronts when measured with the pixels of a camera.
\end{abstract}

\vspace{-2mm}

\end{@twocolumnfalse}
]



\lettrine[lines=3, lhang=0.15]{W}{\:} 
hen light propagates through complex media, such as biological tissue, paint,
clouds or even multimode fibers (MMF), 
it is mixed into a high number of 
degrees of freedom leading to the observation of a 
seemingly random speckle
pattern at the output \cite{rotter2017light,gigan2022roadmap}. 
While the process leading to the generation of this intricate interference pattern
is complex, owing to the deterministic and linear nature of the
propagation of light in such media, the response of the system
between a set of input and output modes is 
fully represented by a single matrix $\bt H$.
This matrix usually corresponds to
the scattering matrix or part 
of it, such as the transmission or reflection 
matrices. While its derivation 
from analytical or numerical models 
is highly challenging and often impossible, 
experimentally, it can be measured 
via wavefront-shaping techniques~\cite{popoff2010measuring,popoff2011controlling}. 
This matrix gives us full knowledge over the wave 
propagation, thus enabling many applications in 
imaging~\cite{popoff2010image,cizmar2012exploiting,gigan2022imaging,gigan2022roadmap}
quantum information~\cite{courme2023manipulation}, 
among many others \cite{rotter2017light,gigan2022roadmap,huepfl2023optimal}.
However, the dynamics of the system or external 
actions introduce perturbations into the known configuration, which 
render our previous knowledge approximate at best. 
For applications in telecommunications and imaging, 
the detrimental effect of this perturbation can be bypassed by finding a set of channels that are 
insensitive to it \cite{mididoddi2023threading,
fan2005principal,carpenter2015observation,ambichl2017focusing,
ambichl2017super,matthes2021learning}. 
When the changes depend on a single parameter $\zeta$ these channels 
can be identified as generalized principal modes which are insensitive 
to first-order variations of $\zeta$ 
\cite{wigner1955lower,fan2005principal,carpenter2015observation,rotter2011generating,
ambichl2017focusing,matthes2021learning,Bouchet2021Maximum}.

Nonetheless, in certain scenarios, the objective may shift from mitigating 
the impacts of perturbations towards actively amplifying
them.
This is the case, for example, when we want to use the output light 
for sensing applications \cite{Bouchet2021Maximum,hill1997fiber,juarez2005distributed,
redding2013all,lealjunior2020optical,gupta2020deep,
cabral2020multimode}. 
By enhancing the interaction between the 
propagating light and the perturbation, the output field becomes 
more sensitive to the perturbation and 
thus carries more information about it, usually 
quantified by the Fisher information $\mathcal J$. 
This increase in information allows 
for increasing the precision when estimating small changes in $\zeta$,
according to the Cram\'er-Rao bound,
which states that the variance of the estimation
$\sigma_\zeta^2$ will be larger or equal to the reciprocal 
of the Fisher information, i.e. 
$\sigma_\zeta^2 \geq \mathcal J^{-1} $
\cite{sengupta1995fundamentals,refregier2004noise}. 
When one uses an external reference to access both the amplitude 
and phase of the output field,
the channel maximizing the Fisher information 
carries almost all the information in its global phase~\cite{Bouchet2021Maximum}.
This presents a significant constraint, 
since phase measurements in optics are highly susceptible to noise, and
require a level of stability usually only available in laboratory conditions, 
making such an approach less suitable for real-life implementations.
In comparison, protocols based on intensity measurements are quite robust 
and therefore broadly applicable 
\cite{redding2013all,lealjunior2020optical,gupta2020deep,cabral2020multimode}.
In particular, the spatial information concealed within the speckle pattern of the light 
coming out of a complex medium has been exploited to develop a wide range
of specklegram sensing devices 
\cite{redding2013all,lealjunior2020optical,gupta2020deep,cabral2020multimode}.
However, the identification of the channel maximizing the 
information carried by its output intensity distribution, and the derivation of the strategy 
that allows reaching the precision limit remain unsolved. 
This is in part due to the fact that 
the relation between the input field and the output 
intensity distribution is not linear, but rather quadratic which prevents the 
use of standard matrix methods. 

\begin{figure*}[t]
  \centering
  \includegraphics[width=.9\linewidth]{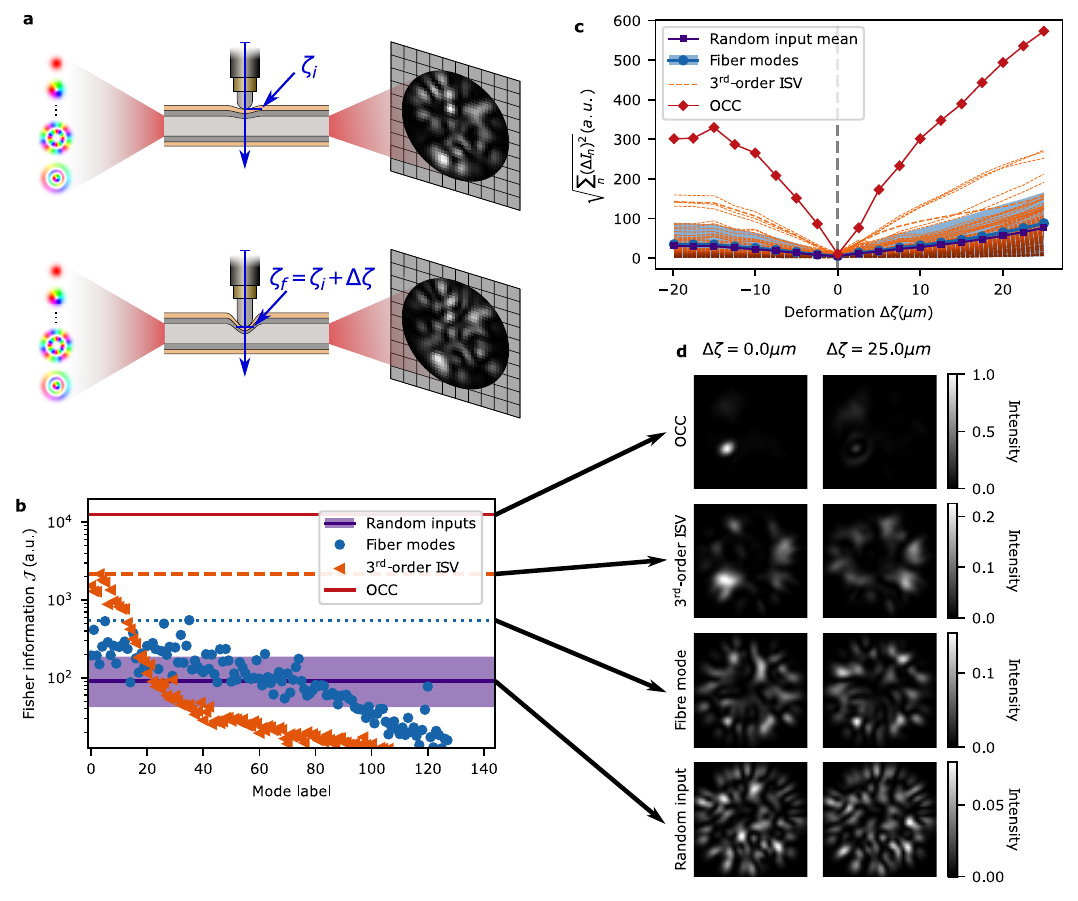}
  \figurecaption{%
  Sensitivity to perturbations%
  }{%
   \bt a, Change on the output intensity distribution 
   induced by a local deformation on a MMF. 
   \bt b, 
   Fisher information in the pixel basis for random inputs (where
   the line denotes the mean value for one-thousand states and 
   the shaded region the range obtained), the fiber modes (where the 
   dotted line marks the maximum value), the third-order input singular vectors (ISVs) (where 
   the dashed line marks the maximum value), and the optimal crashing channel (OCC).
   \bt c,
   Intensity change over large deformations for the same modes 
   as in (\bt b). For the random inputs only the mean is shown, and for the fiber
   modes the marked line denotes the mean and the shaded region the range of values.
   \bt d, Output intensity distribution 
   for the OCC, the most sensitive
   third-order ISV, the most sensitive fiber mode, as well as a random input for the 
   reference deformation $\Delta \zeta =0 \mu m$ and a large deformation 
   for which $\Delta \zeta = 25 \mu m$. 
  }
  \label{fig:sum}
\end{figure*}

To solve these problems, we exploit the versatility of higher-order 
tensors to describe the quadratic relationship between the input field 
and its output intensity distribution.
We use this tensor-based framework to study 
three practical configurations for robust intensity-based sensing 
applications.
First, when we only have control over the input wavefront for a 
fixed detection scheme 
at the distal end of the fiber. 
Second, when controlling the projection of the outgoing field 
on the distal end for a fixed arbitrary input. 
Third, when controlling both the input wavefront and the output projections.
In particular, we experimentally study the case 
of an MMF which is perturbed 
by pressing down on it transversally with a motorized actuator
as exemplified in Fig.~\sref{fig:sum}{a}.
The parameter $\zeta$ represents
the linear displacement of the actuator and $\bt H$ the 
transmission matrix (TM) between the input modes 
formed by the 144 
modes of the fiber (see Methods for more details), 
and a set of output modes, such 
as the pixels of a camera.
With this implementation we are able to demonstrate 
an enhancement in the Fisher information 
by more than four orders of magnitude when using 
the optimal input-output configuration.

\section{Results}
\subsection{Optimizing the input field}
Intuitively, the channel maximizing the Fisher information 
carried by its output intensity distribution, under the assumption 
of Gaussian noise with a known standard deviation $\sigma$,
is the one for which first-order variations in $\zeta$ 
lead to the largest change in its output intensity distribution. 
Therefore, this channel can be seen as crashing straight 
into the perturbation, and thus we refer to it as the 
\emph{optimal crashing channel} (OCC).
The assumption of Gaussian noise is quite general encompassing
all systems for which the noise fluctuations are dominated by dark and readout noise.

Given the quadratic dependence of the intensity on the field, one may think 
that there is no other choice than to cast this problem as a standard
nonlinear optimization problem~\cite{ambichl2017super,bouchet2020influence,bouchet2021optimizing}. 
Nevertheless, higher-order tensors,
which are multidimensional extensions of matrices,
provide fresh perspectives in tackling intricate and nonlinear 
challenges~\cite{tucker1966some,lathauwer2000best,
Lathauwer2000Multilinear,
comon2009tensor,kolda2009tensor,kolda2014adaptive,
sidiropoulos2017tensor,rabanser2017introduction}. 
In fact, as shown in the Methods, for a fixed set of output modes 
(such as the pixels of a camera)
it is possible to rewrite the Fisher information as
\begin{equation} \label{eq:innerw3}
  \mathcal{J}(\zeta) =   \frac{1}{\sigma^2} 
  \langle  \ten{W}^{(3)}, \bt a \otimes
  \bEin  \otimes \bEin{}^* \rangle^2,
\end{equation}
which is the inner product between the third-order tensor $\ten W^{(3)}$, 
defined component-wise as 
$\mathcal{W}^{(3)}_{ijk} = \partial_\zeta ( \text{H}_{ij}^*\text{H}_{ik})$ where 
$\text{H}_{ij}$ are the components of the matrix $\bt H$,
and the rank-one third-order tensor $\bt a \otimes
\bEin  \otimes \bEin{}^*$ where $\bt a$ is a normalized real vector 
quantifying the change in the output intensity of the input field $\bEin$ over the
output modes and $\otimes$ denotes the outer 
product (see Methods for more details). Maximizing $\mathcal{J}$ for a fixed amount 
of total energy at the input 
is equivalent to finding the best rank-one approximation of 
$\ten W^{(3)}$, that is, the set of three vectors that best approximate
it \cite{kolda2009tensor,sidiropoulos2017tensor}.

\begin{figure}[t]
  \centering
  \includegraphics[width=.99\linewidth]{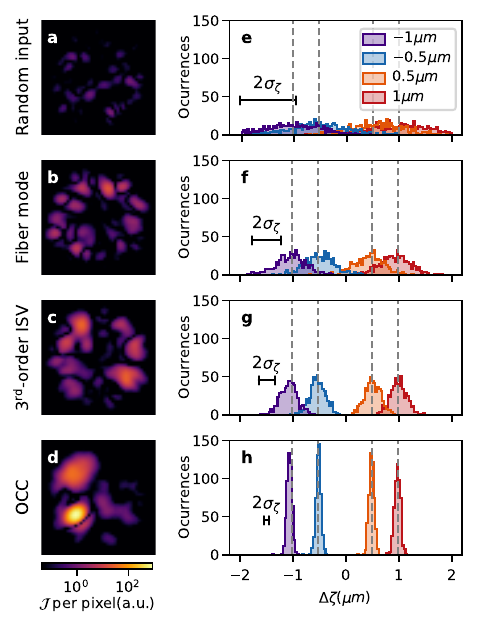}
  \figurecaption{Estimating a change in the perturbation}{ \bt a-\bt d
  Measured Fisher information per output pixel.
  \bt e-\bt h Histograms of estimated changes in 
  the deformation for four different values 
  ($\Delta\zeta = -1,-0.5,0.5,1 \mu m$) marked with dashed gray lines and 
  with the mean width $2\sigma_\zeta$ marked in black.
  Both types of plots are shown for a random input (\bt a, \bt e), 
  the most sensitive fiber mode (\bt b, \bt f), the optimal third-order 
  ISV 
  (\bt c, \bt g),
  and the OCC  (\bt d, \bt h).}
  \label{fig:estim} 
\end{figure}

For matrices, which are second-order tensors, 
finding the best rank-one 
approximation is a simple task achieved by computing 
the singular-value decomposition (SVD) and taking the outer product of 
the first pair of
singular vectors. For higher-order tensors, however, 
the answer is not so simple. Nonetheless, a
step in the right direction can be taken by performing the higher-order
SVD (HOSVD)
\cite{Lathauwer2000Multilinear,comon2009tensor,kolda2009tensor}. 
This generalization of the matrix SVD allows decomposing 
a higher-order tensor in terms of a sum of rank-one tensors 
composed of higher-order singular vectors which form orthonormal bases for
their respective spaces. 

To find the third-order input singular vectors
(ISVs) of $\ten W^{(3)}$ we first construct this tensor from two measurements of $\bt H$, using the pixels of a camera as output modes, 
around a reference value $\zeta = \zeta_i$
for the deformation, and then compute its HOSVD.
In Fig.~\sref{fig:sum}{b} we compare the values of the Fisher information obtained for the third-order ISV with those obtained when using 
the fiber modes and one-thousand random wavefronts as inputs.
It can be seen that the Fisher information for the first third-order ISVs is above all
the values obtained with the fiber modes and random wavefronts, and with 
the maximum value being an order of 
magnitude larger the maximum value attained for a random wavefront. 
Therefore, the HOSVD immediately provides us with a set of modes generally ordered 
from the one with the largest Fisher information to the smallest, 
and with the first ones being 
highly sensitive to the perturbation.


\emph{Optimal crashing channel.} One of the main differences between the HOSVD 
and the SVD is that the first singular vectors do not immediately provide us
with the best rank-one approximation.
Nonetheless, they form a reasonably good first guess that can
be used as a seed for a modified version of the iterative
alternating-least squares algorithm~\cite{kolda2009tensor,sidiropoulos2017tensor}
to determine the best rank-one approximation (see Methods for details).
This allows us to identify the OCC which
provides a two order of magnitude boost in Fisher information 
with respect to an average random wavefront, as 
shown in Fig.~\sref{fig:sum}{b}. 

It is also worth noting that, as seen in Fig.~\sref{fig:sum}{c}, 
both the third-order ISVs and the OCC maintain
their high sensitivity over a large range of 
deformations. However, by looking at the intensity distributions
in Fig.~\sref{fig:sum}{d}, it can be seen that their behavior is
quite different. While the intensity distribution for the most sensitive
third-order ISV changes radically, for the OCC it is mostly the total intensity
that suffers the largest change and not the overall distribution.
This feature of the OCC is a consequence of the 
non-unitarity of the matrix $\bt H$ for MMF which can be 
potentially very useful for practical implementations. Indeed, 
such a change can be recorded by a single detector, 
thus minimizing the cost of the system
and allowing high-speed operations.
Another feature worth noting
is that the intensity distribution of the OCC is mostly concentrated around
a focal spot. This allows it to focus as 
much information across a few pixels in order 
to increase the signal-to-noise ratio (SNR). 
Despite the similarity, the OCC is not equivalent to 
the channel obtained through phase 
conjugation to focus at the same spot \cite{popoff2011controlling}
and for which the Fisher information value is much lower. 

\emph{Estimating the perturbation.} A direct application 
of the OCC regards the estimation of changes in the perturbation  
$\Delta\zeta$. 
Assuming $\Delta \zeta$ to be small, when compared to the deformation
that decorrelates the output, we can safely assume a linear model
for which we only need to perform calibration measurements for 
the reference intensity distribution and its derivative with respect to 
$\zeta$ above the noise level. 
Figures \sref{fig:estim}{a-d} show
the Fisher information per output pixel obtained from these calibration measurements for 
four different input fields, 
namely, a random wavefront, the fiber mode with the highest Fisher information, 
the third-order ISV with the highest Fisher information
and the OCC. Here, the drastic enhancement provided by the OCC can be appreciated.
For the actual estimation, we use these same four fields but perform the 
measurements at a lower input power and for the four smaller deformation values 
$\Delta \zeta = -1,-0.5,0.5, \text{ and } 1 \mu m$. 
We measure 500 intensity distributions for each field 
and each deformation, 
and use each one of these measurements to estimate the deformation
(see Methods for more details). 
The results are shown in Figs.~\sref{fig:estim}{e-h}. For the 
random input, it is practically impossible to discriminate 
the four peaks corresponding to the four different deformation values. 
For the fiber mode and the third-order ISVs, we can clearly discern
the deformations that are further apart as the distributions are narrower. 
However, there is still 
significant overlap between the ones that are closer together. 
Finally, when we take a look at the results for the OCC, we can 
appreciate four well-defined peaks with a standard deviation 
that is more than an order of magnitude smaller 
than the one obtained for the random input, in agreement with 
the Cram\'er-Rao lower bound. 

\subsection{Optimizing the output projection modes}
One consequence of the quadratic relationship between the input field and the 
measured output intensity is that
the Fisher information is not invariant under 
changes of the output modes. 
While the pixels of a camera might be the simplest output modes 
to implement experimentally,
they are generally not the optimal choice since they are
blind to the information that could 
be hidden in the relative phase
variations of the field from one pixel to another, and 
they spread out the information 
across many modes, thus decreasing the SNR.
To address both of these issues we could foresee demultiplexing the output 
field into a specifically designed set of spatial channels, 
which is experimentally feasible using photonic 
lanterns or multiplane light converters 
\cite{fontaine2012geometric,labroille2014efficient,fontaine2019laguerre}. 
This is the principle behind techniques that allow  
increasing the information 
transfer for telecommunication applications 
\cite{bozinovic2013terabit,cristiani2022roadmap} or beating Rayleigh's curse 
when imaging two closely-spaced sources
\cite{tsang2016quantum,paur2016achieving,liang2021coherence,treps2002surpassing}.

\begin{figure}[t]
  \centering
  \includegraphics[width=.99\linewidth]{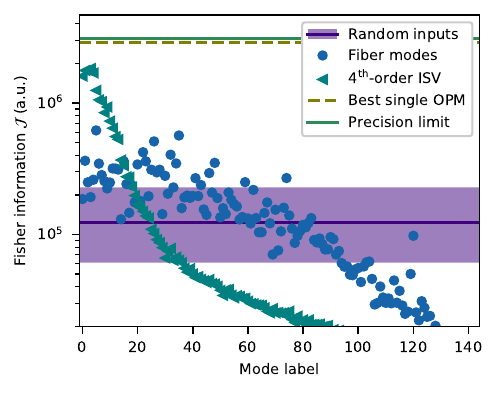}
  \figurecaption{Reaching the precision limit}{ 
  Effect of using the optimal output projection modes (OPMs)
  on the Fisher information per photon 
  for various input fields. It shows the Fisher information
  obtained when using the optimal OPMs for 
  each of the random inputs used in Fig.~\ref{fig:sum}, the fiber
  modes, and the fourth-order ISVs. Also shown 
  are the Fisher information values attained for the optimal input-output combination, which 
  defines the precision limit for intensity-based measurements, 
  and for the optimal channel when a single OPM is being used at the output.
  }
  \label{fig:fisher_moim}
\end{figure}

Mathematically, this spatial demultiplexing is performed by projecting
the output field onto an orthogonal set of $Q$ output projection modes (OPMs),
where $Q$ can smaller than the number of output modes used to define $\bt H$.
If we assume that the input field is fixed, then, as shown in the Methods,
the Fisher information can be rewritten as 
\begin{align}
  \mathcal{J}(\zeta) 
    = & \frac{1}{\sigma^2} \sum_{q=1}^Q \langle  \bt p_{q} , 
     \bt E_\zeta 
    \cdot \bt p_{q} \rangle^2,
\end{align}
where $\bt p_q$ is the $q^{\text{th}}$ OPM, and
$\bt E_\zeta = \partial_\zeta \left(\bEout \otimes \bEout{}^* \right)$
is a rank-2 Hermitian matrix. Therefore, the Fisher 
information can be maximized
by choosing as OPMs the two eigenvectors of $\bt E_\zeta$ 
with nonzero eigenvalues.
These two optimal OPMs are given by a simple
linear combination of the output field $\bEout$ and its derivative with 
respect to the parameter $\partial_\zeta \bEout$. This is similar to 
the approaches taken for estimating the distance between two particles 
where the output field is projected on a Gaussian field, which resembles the 
point spread function, and higher-order 
Hermite-Gauss modes, which provide the derivatives 
\cite{tsang2016quantum,paur2016achieving,liang2021coherence}. 

Figure \ref{fig:fisher_moim} shows the impact of projecting onto the
optimal OPMs on the Fisher information
for the same random inputs and fiber modes as those used in 
Fig.~\sref{fig:sum}{b}. 
By using the optimal OPMs, the mean Fisher information obtained
when using random wavefronts as inputs is an order of magnitude larger 
than the maximum value obtained with the OCC in the pixel basis,
which clearly shows the benefits 
of choosing the OPMs appropriately 
even when one cannot control the input wavefront.

\subsection{Optimal input-output combination: reaching the precision limit}
Let us now find the optimal channel when we are free to shape the input field
and choose the OPMs. This optimal input-output combination is the channel that
maximizes the Fisher information when projected onto its optimal OPMs, and
sets the precision limit achievable with intensity measurements.
To find it, we rewrite the Fisher information in terms
of just two OPMs and the input field which leads to,
\begin{align} \label{eq:fishw4}
  \mathcal{J}(\zeta) 
    = & \frac{1}{\sigma^2} \sum_{q=1}^2 
    \langle \ten{W}^{(4)}, \bt p_q^* \otimes \bt p_q \otimes 
    \bEin \otimes \bEin{}^*
    \rangle^2,
\end{align}
where we defined the fourth-order tensor $\ten{W}^{(4)}$ whose 
components are given by 
$W^{(4)}_{ijkl} =  \partial_\zeta (\text{H}_{ik}^*\text{H}_{jl})$.
Even though this expression resembles the one in Eq.~(\ref{eq:innerw3}),
it does not correspond to a rank-two approximation of $\ten{W}^{(4)}$.
Nonetheless, we can still use tensor-based techniques
to obtain an excellent first guess by computing the HOSVD of $\ten{W}^{(4)}$.
Figure~\ref{fig:fisher_moim} shows the resulting Fisher information
for the corresponding fourth-order ISVs 
when projected onto their respective
optimal OPMs. Once more, these higher-order ISVs 
provide us with an ordered orthogonal basis of highly-sensitive modes,
with the first ones surpassing all the modes of the fibers and random inputs.

\begin{figure}[t]
  \centering
  \includegraphics[width=.99\linewidth]{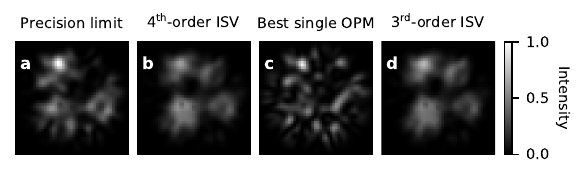}
  \figurecaption{Output fields near the precision limit}{ 
  Output 
  intensity distribution for the mode achieving the precision 
  limit
  (\bt a), 
  the first fourth-order ISV
  (\bt b), 
  the first third-order ISV
  (\bt c), 
  and the best mode when using a single OPM (\bt d).
  }
  \label{fig:comp_fields}
\end{figure}

To reach the precision limit, however, we need to perform a nonlinear optimization using 
the first fourth-order ISV as a seed (see Methods for details). 
The results shown in Fig.~\ref{fig:fisher_moim} demonstrate that the Fisher information 
achieved by this channel is well over two orders of magnitude above that obtained for 
the OCC that uses the pixels as OPMs. 
Nevertheless, this maximum value is very close to that provided
by the first few fourth-order ISVs.
Another approach consists on taking the first 
fourth-order ISV as a seed to find the 
best rank-one approximation of $\ten W^{(4)}$. While this does not allow 
us to reach the precision limit, it provides us with the optimal mode 
when a single OPM is used to monitor the change of intensity at the output, 
that is when there is a single term in 
Eq.~(\ref{eq:fishw4}). As can be seen in 
Fig.~\sref{fig:fisher_moim}, this almost allows us to reach the precision limit, for which
the Fisher information value is only 4.8\% larger, while using a single OPM. 
For practical applications, it means that one can approach the precision 
limit within a very small margin using a single photodetector.
In fact, the output fields produced by the first fourth-order ISV and 
the best rank-one approximation are highly similar to the one reaching 
the precision limit, with which they have 
field correlations that are above 78\%.
This similarity can also be appreciated 
in their intensity profiles, shown in Fig.~\ref{fig:comp_fields}.
More surprising, however, is that the first third-order ISV obtained 
with the output pixel basis is also highly similar to the one reaching the precision
limit (see Fig.~\ref{fig:comp_fields}). This is not intuitive since 
$\ten W^{(3)}$ does not contain any information about the phase at the output, 
which $\ten W^{(4)}$ does have.

\section{Outlook}

By introducing tensor-based methods to the study of complex systems, 
we have provided a natural framework for 
studying intensity-based measurements.
Here, we used lower-rank approximations of higher-order tensors to find the channels 
that are most sensitive to a given perturbation i.~e.~that suffer the largest
change in output intensity when the perturbation changes.
Their high sensitivity allows them to be used for highly robust and precise
sensing applications which we demonstrated experimentally 
by estimating small perturbations in MMF.
It was also shown that what is meant by most sensitive is highly 
dependent on the choice of output modes used to measure the intensity 
distribution. This dependence was 
exploited to find the channel that allows extracting the most information 
available at the output, and thus achieve the precision limit of
intensity-based measurements.

This new framework opens the door to further investigations, 
such as the identification of universal input or output modes
for blind parameter estimation, that is when the 
output modes or the input field are not known or can change over time. 
Additionally, the tensor-based relations established here can also be used 
for the 
study of incoherent systems where a TM cannot be derived. 
This last point could help gain further insight into fluorescent 
imaging applications through scattering media. 
Likewise, it can be expected for the fields presented here to 
find other applications, such as for focusing light inside scattering 
media but with properties 
that will be quite different when using the generalized principal 
modes~\cite{ambichl2017focusing}. 






\bibliographystyle{arthur} 
\bibliography{refs}{}

\newpage
 
\section{Methods}
\subsection{Fisher information for intensity measurements.} 
Given a random process determined by a probability mass function 
$P(\bs \chi|\zeta)$ for the set of 
random variables $\bs \chi$ and with parametric dependence on $\zeta$, the 
Fisher information for the parameter $\zeta$ is defined as
\begin{align}
  \mathcal{J}(\zeta) = \overline{ \left[ \partial_\zeta 
      \ln P(\bs \chi|\zeta)\right]^2},
\end{align}
where the overline denotes the 
expected values over all the possible values of $\bs \chi$.
Here, $P(\bs \chi|\zeta)$ represents the 
inaccuracy of the measured the intensity distribution over the $N$ output modes
$\{ \bt y_1, \ldots, \bt y_N \} $ due to the presence of
noise. Thus, the measured intensity distribution is given by 
$\bs \chi = \bt I + \bs{\mathscr{w}}$ where $\bt I = |\bEout|^2$ is the expected 
intensity distribution and $\bs{\mathscr{w}}$ represents the noise which is assumed to 
be Gaussian with known standard deviation $\sigma_n$ 
for the n${}^\text{th}$ output mode. The probability mass function 
can then be written as
\begin{align}
  P(\bs \chi|\zeta) = \prod_{n=1}^N  \frac{1}{\sigma_n \sqrt{2\pi}}
  e^{- \frac{(\chi_n - I_n)^2}{2\sigma_n^2}},
\end{align}
The intensity value of the n${}^\text{th}$ 
output mode is defined as $I_n=|\Eout_n|^2$ where $\Eout_n$ is the coefficient 
of the n\textsuperscript{th} output mode
in the linear expansion $\bEout = \sum_n \Eout_n \bt y_n$.
The Fisher information then takes the following
simple form
\begin{align}
  \mathcal{J}(\zeta) = \sum_{n=1}^N \frac{1}{\sigma_n^2}
  \left[ \partial_\zeta |\Eout_n|^2 \right]^2.
\end{align}

\emph{Fisher information as a rank-one approximation.} 
Writing the output field in terms of the matrix $\bt H$ and the input 
field, which is decomposed in terms of the input modes 
$\{ \bt x_1, \ldots, \bt x_M \} $, 
$\bEin = \sum_{m=1} \Ein_m \bt x_m$, and normalized, $||\bEin ||=1$, 
we get that 
\bse
\begin{align}
  \mathcal{J}(\zeta) = &\frac{1}{\sigma^2}\sum_{n=1}^N 
  \left[ \sum_{m,m'}^M\partial_\zeta (\text{H}_{nm}\text{H}_{nm'}^*) 
  \Ein_m\Ein_{m'}{}^*\right]^2\\
  =& \frac{ \mathcal{J}^{1/2}(\zeta)}{\sigma}\sum_{n=1}^N \sum_{m,m'}^M  
   W_{nmm'}^* a_n \Ein_m\Ein_{m'}{}^*, \\
  =& \frac{ \mathcal{J}^{1/2}(\zeta)}{\sigma}\left\langle \ten W^{(3)}, 
  \bt a \otimes \bEin \otimes \bEin{}^* \right\rangle,  \label{eq:devfish}
\end{align}
\ese
where the real vector $\bt a$ is defined component-wise as
\begin{align}
  a_n = \frac{1}{\sigma \mathcal{J}^{1/2}(\zeta)}\sum_{m,m'}^M
   W_{nmm'} \Ein_m\Ein_{m'}{}^*,
\end{align}
with $\norm{\bt a} = \mathcal{J}^{1/2}(\zeta)$ so that it is normalized.
This real vector $\bt a$ is proportional to the rate of change of the intensity
distribution over the output modes with respect to the parameter $\zeta$ for the input 
field $\bEin$. Its value is fixed as the one that maximizes $\mathcal{J}$ 
for a fixed $\bEin$.
For simplicity, we have taken the noise over the output modes to be
the same, $\sigma = \sigma_n$ for all $n$, but, if needed, these 
output mode-dependent noise variations can be incorporated in the definition of 
$\ten W^{(3)}$.
Finally, by squaring both sides of Eq.~\ref{eq:devfish}, 
the Fisher information can be rewritten as 
\begin{align}
  \mathcal{J}(\zeta) = \frac{1}{\sigma^2} \left\langle \ten W^{(3)}, 
  \bt a \otimes \bEin \otimes \bEin{}^*\right\rangle^2.
\end{align}
Maximizing $\mathcal{J}$ subject to the constraints 
$\norm{\bt a}=||\bEin ||=1$ is equivalent to the minimization problem
\begin{align} \label{eq:ssroa}
  \min_{\bt u, \bt v, \bt w}\norm{\ten W^{(3)} - \bt u \otimes \bt v \otimes \bt w},
  \text{ subject to } \frac{\bt v}{\norm{\bt v}} = \frac{\bt w^*}{\norm{\bt w}} ,
\end{align}
where $\bt u$, $\bt v$ and $\bt w$ are vectors, thus showing that 
finding the OCC 
is equivalent to finding the best rank-one approximation 
of $\ten W^{(3)}$ \cite{lathauwer2000best}.
The OCC $\bEin_\text{OCC}$ can then be identified from the solution as
\begin{align}
  \bEin_\text{OCC} = \bt v / \norm{\bt v}.
\end{align}

\emph{Changing the OPM.} 
To project the output field onto a set of $Q$ orthonormal OPM, 
we simply apply the corresponding projection operator
$\bt P$ to the output field. 
The projection operator satisfies  $\bt P^\dagger \cdot 
\bt P = \mathds{1}_Q$ where $\mathds{1}_Q$ is the identity matrix of dimension $Q$.
Note that $\bt P$ is generally not unitary since $Q$ can be less than the 
total number of output modes in the system. 
The projected output field can then be written as 
\begin{align}
  \bEout_P = \bt P^\dagger \cdot \bEout 
  = \sum_{q=1}^Q \Eout_{P,q} \bt p_q,
\end{align}
where $\bt p_q$ is the $q^{\text{th}}$ column of $\bt P$, i.e. the 
$q^{\text{th}}$ OPM, and the intensity distribution over 
the new OPM 
is given by $I^{(P)}_q= |\Eout_{P,q}|^2$.
Rewriting the Fisher information in terms of an output modes basis that contains 
all the information about the output field we have that,
\bse
\begin{align}
  \mathcal{J}(\zeta) 
  = & \frac{1}{\sigma^2} \sum_{q=1}^Q 
  \left[  
  \partial_\zeta|\Eout_{P,q}|^2
    \right]^2 , \\
  = & \frac{1}{\sigma^2} \sum_{q=1}^Q 
  \left[  \sum_{n,n'} P_{nq}^*
  \partial_\zeta(\Eout_n \Eout_{n'}{}^*) P_{n'q}
    \right]^2 , \label{eq:pfish} \\
    = & \frac{1}{\sigma^2} \sum_{q=1}^Q \langle  \bt p_{q} , 
     \bt E_\zeta 
    \cdot \bt p_{q} \rangle^2
\end{align}\
\ese
where 
\begin{align}
  \bt E_\zeta = 
    \partial_\zeta \left( \bEout \otimes \bEout{}^* \right).
\end{align}
It is easy to see that the $\bt E_\zeta$ operator is Hermitian
and of rank two. Therefore, it only has two non-zero eigenvalues, 
which means that all the information available in $\bEout$ can
be obtained by projecting into two modes. Note that this result
applies even to changes happening at the source and not only those
encoded in the matrix $\bt H$.

\emph{Fisher information for optimal input-output combinations.}
Now, if we want to find the optimal input-output combination in order to 
reach the precision limit for intensity-based measurements, then we need
to write the Fisher information in terms of the input 
field. Moreover, as we just showed, we only need to consider projection 
the output into two OPM. Therefore, setting $\bEout = \bt H\cdot \bEin$ and
$Q=2$ in Eq.~(\ref{eq:pfish}), we get
\bse
\begin{align}
  \mathcal{J}(\zeta) 
  = &\frac{1}{\sigma^2} \sum_{q=1}^Q 
  \left[  \sum_{n,n'} \sum_{mm'}\partial_\zeta(\text{H}_{nm} \text{H}_{n'm'}^*)
    P_{nq}^* P_{n'q}\Ein_m \Ein_{m'}{}^*
  \right]^2 ,  \\
  = & \frac{1}{\sigma^2} \sum_{q=1}^Q\langle 
    \ten{W}^{(4)}, \bt p_q^* \otimes \bt p_q \otimes 
    \bEin \otimes \bEin{}^*
    \rangle^2,
\end{align}\
\ese
where we defined the fourth-order tensor $\ten{W}^{(4)}$ whose 
components are given by
\begin{align}
  W^{(4)}_{ijkl} =  \partial_\zeta (\text{H}_{ik}^*\text{H}_{jl}).
\end{align}


\subsection{Experimental setup}
The optical setup is represented in \refedfig{fig:setup}.
The light source consists of a continuous linearly 
polarized laser beam at 1550 nm 
(TeraXion NLL) injected into a 10:90 polarization-maintaining fiber 
coupler (PNH1550R2F1) in order to separate it into the shaped 
signal field and the reference field. 
The 90\% arm is collimated and expanded to 
illuminate a digital micromirror device (DMD) (Vialux V-650L) 
which is used to modulate the signal field in amplitude
and phase using Lee holograms \citesupp{lee1978iii}. The light is converted into left 
circular polarization using a quarter-wave plate. The shaped signal 
field is then imaged with a 4f system onto the input facet of 
a 25cm-long step-index fiber with a 50 $\mu m$ core and 0.22 
numerical aperture, which is held
approximately straight. 
The output facet is imaged via another 4f system onto an InGaAs camera
(Xenics Cheetah 640-CL 400 Hz) after passing through a quarter-wave 
plate, followed by a beam displacer
to select the left-circularly polarized component.
The other 10\% arm is used to produce a tilted reference that is made 
to interfere with the signal field in order to be able to retrieve the
output field via off-axis holography \citesupp{cuche2000spatial}. 
An Arduino-controlled shutter
allows blocking the reference field to perform intensity measurements
of the signal field. 
The fiber can be deformed, roughly in the middle, by pressing on it using 
a 50 nm precision dc servo motor actuator (Thorlabs Z812).
All the values reported for the deformation correspond to the linear 
displacement of the actuator and not the deformation of 
the fiber core, which is much smaller since most of the deformation 
is absorbed by the coating of the fiber. 
\\

\subsection{TM measurements} First, we measure the TM of the MMF in the pixel 
basis, by sending 7200 square layouts consisting of $37\times37$ square macropixels
whose value is either zero or a random phase of amplitude one. Each 
macropixel is formed by grouping $16\times16$ pixels of the DMD, where the 
desired phase and amplitude value is encoded via
Lee holograms \citesupp{lee1978iii} and selecting the first order of diffraction 
at the Fourier plane. The corresponding output fields are recovered 
from the interferograms between the reference and signal fields, and subsequently projected onto a square pattern of $44\times44$
macropixels formed by grouping $4\times4$ pixels of the camera.
Regrouping all input and output fields into the columns of matrices $\bt X$
and $\bt Y$, respectively, we reconstruct the TM via 
$\bt H = \bt Y \cdot \bt X^{-1}$ where $\bt X^{-1}$ denotes the pseudoinverse 
of $\bt X$.

Then, we compute the SVD of the TM in the pixel 
basis and use it to identify the fiber modes as the singular vectors with
singular values in the almost constant plateau (see \refedfig{fig:svdpix}). 
For the fiber we used there
were 144 fiber modes as opposed to the 129 predicted 
through numerical simulations. This indicates that some cladding modes 
are sufficiently close to the cutoff that they propagate 
without undergoing significant absorption for the short length we are 
using. 
All subsequent TM measurements are performed by sending 1440 random inputs 
obtained by randomly superimposing all 144 fiber modes. This allows us to 
drastically reduce the size of the TM and the higher-order tensors without 
losing significant information. The output fiber modes are simply the outputs generated by sending the fiber
modes as inputs. These outputs are orthogonal since they correspond to 
the first 144 output singular vectors of the TM measured in the pixel basis.

\subsection{Optimizing the Fisher information}
When the output modes are fixed, the OCC for which the 
Fisher information achieves a maximum can be identified from the best
rank-one approximation of $\ten W^{(3)}$. As mentioned in the main text, 
for matrices which are second-order tensors finding the best rank-one
approximation is quite simple since one only need to compute the SVD. 
For tensors, it is not so simple, and thus we take a few steps to get there. 

\emph{Higher-order singular value decomposition.} 
A first step into the right direction can be taken by computing the HOSVD 
which is defined in what follows.
A tensor $\ten{T} \in \mathbb{C}^{I_1}\times \cdots 
\times \mathbb{C}^{I_L}$ of order $L$
can be decomposed as a sum of rank-one tensors 
using a higher-order version of the SVD,
\begin{align}
  \ten{T} = \sum_{i_1=1}^{I_1}\cdots \sum_{i_L=1}^{I_L} s_{i_1 \cdots i_L}
   \bt u^{(1)}_{i_1} 
  \otimes \cdots \otimes \bt u^{(L)}_{i_L}
\end{align}
where the  higher-order singular vectors $\bt u^{(j)}_i$ form 
an orthonormal basis 
of $\mathbb{C}^{I_j}$,
$S_{ijk}$ are the entries of the core tensor $\ten S$, and $\otimes$
denotes the outer product. The core tensor satisfies the properties of 
\begin{enumerate}
  \item all orthogonality:
  $\langle \ten{S}_{i_n=\alpha},\ten{S}_{i_n=\beta} \rangle = 0$ 
  for $\alpha \neq \beta$
  \item ordering: $\norm{\ten{S}_{i_n=1}}\geq\norm{\ten{S}_{i_n=2}}
    \geq \ldots$,
\end{enumerate}
where $\langle \cdots , \cdots \rangle$ denotes the inner 
product induced by the Frobenius norm, namely 
\begin{align}
  \langle \ten{T}^{(1)} , \ten{T}^{(2)} \rangle = 
  \sum_{i_1=1}^{I_1}\cdots \sum_{i_L=1}^{I_L} T^{(1)\;*}_{i_1 \cdots i_L}{}
  T^{(2)}_{i_1 \cdots i_L},
\end{align}
and $\ten{S}_{i_n=\alpha}$ denotes the tensor obtained by setting
the $n^{\text{th}}$ index equal to $\alpha$.
For matrices, these properties reduce to the diagonality condition 
of the matrix made of singular values and their ordering.

Another important property of the HOSVD is that it preserves 
the partial symmetries of the original tensor, 
for example if $T_{i_1 \cdots i_j \cdots i_k \cdots i_L}
= T^*_{i_1 \cdots i_k \cdots i_j \cdots i_L}$ then 
$\bt u^{(j)}_{i} = \bt u^{(k)}_{i}{}^*$ and 
$S_{i_1 \cdots i_j \cdots i_k \cdots i_L}
= S^*_{i_1 \cdots i_k \cdots i_j \cdots i_L}$.
The biggest advantage of the HOSVD against other types of tensor
decompositions is that it can be easily computed via the SVD of 
its matrix unfoldings (see \citers{Lathauwer2000Multilinear,kolda2009tensor} 
for more details). However, when truncating the HOSVD to a single term we do 
not get the best rank one approximation, but we do get an excellent first guess.

\emph{Symmetrized alternating least-squares algorithm.}
Finding the best rank-one approximation is a particular case of more general
types of lower-rank tensor approximations based around the polyadic 
and Tucker decompositions \cite{comon2009tensor,kolda2014adaptive}. There are several algorithms that have been 
developed to find them and that could, in principle, 
be applied to solve the optimization problem in Eq.~(\ref{eq:ssroa}). 
However, most of them 
do not take into account the symmetries of the original tensor, and when 
one tries to impose them they often fail to converge. 

Here, we were able to adapt the widely used alternating-least squares (ALS)
algorithm in order to keep the symmetries of the original tensor. 
Let us consider the case of a third-order tensor $\ten{T}$ 
with partial Hermitian 
symmetry in the last two indices, $T_{ijk}= T_{ikj}^*$, where we seek
the vectors $\bt a$ and $\bt b$ that solve
\begin{align}
\min_{\bt a, \bt b}\norm{\ten T - \bt a \otimes \bt b \otimes \bt b^*}.
\end{align}
Initially, 
we do not force the symmetry and allow the last vector to be different, 
say $\bt c$ and start by minimizing 
$\norm{\ten T - \bt a \otimes \bt b \otimes \bt c}$. We proceed 
as a standard ALS routine by fixing $\bt b$ and $\bt c$, and solving the 
minimization problem
\begin{align}
  \min_{\bt a} \norm{\ten T - \bt a \otimes \bt b \otimes \bt c},
\end{align}
which is a standard least-squares problem with a fixed solution.
Then, we do the same for $\bt b$ and $\bt c$. Before performing 
another loop, we symmetrize the lower rank tensor by setting
\bse
\begin{align}
  \bt b \gets & \frac{\sqrt{\norm{\bt b}\norm{\bt c}}}{2} 
    \left(\frac{\bt b }{\norm{\bt b}} + \frac{\bt c^* }{\norm{\bt c}} \right) \\
  \bt c \gets & \bt b^*
\end{align} 
\ese
This simple adaptation gives enough freedom so the algorithm 
converges and provides a solution with the desired symmetry. 
A similar approach is taken to compute the rank-one approximation 
of $\ten W^{(4)}$.

The ALS algorithm requires us to provide it with a starting point, that is 
a first guess for the best rank-one approximation. Here, we use as a seed
the rank-one tensor obtained by truncating the HOSVD to a single term. 
This choice allows us to systematically converge to the optimal solution 
whereas when using a random guess there is a high probability to fall into 
a local minimum.

\emph{Finding the optimal input-output combination.} The first step in finding 
the optimal input-output combination is to compute the fourth-order tensor
$\ten W^{(4)}$. Given the number of pixels used as output modes to measure $\bt H$,
the size of this tensor would be too large to be useful for computations. 
However, since the number of input modes is much lower, one can see that using 
all the 1444 output pixels might be an overkill. 
Therefore, we first compute the SVD for the 
two TMs that we used to compute $\ten W^{(3)}$ and use their output 
singular vectors to form an orthonormal 
set of output modes. This new set of output modes has double 
the number of input 
modes used, but it allows us to capture all the information encoded in 
the pixel basis but at a much smaller size. 
Then, we project the TMs onto this set of output modes  
and use this projected TMs to compute $\ten W^{(4)}$.

As mentioned in the main text, 
when trying to find the optimal input-output combination that
allows reaching the precision limit, the Fisher information given by 
Eq.~(\ref{eq:fishw4}) no longer
takes the form of a lower-rank tensor decomposition. Nonetheless, it is 
possible to rewrite it as a nonlinear optimization where only the expansion 
coefficients of the input field in terms of the input modes are used as 
optimization parameters. 
This is done by first computing the output field using the TM, then 
getting the optimal OPMs and using those to compute the Fisher information. 
This optimization was implemented using the neural network framework
PyTorch. Here again, we need to provide a first guess, and, as for the ALS algorithm, 
we chose to use the first term of the HOSVD of $\ten W^{(4)}$. 

\subsection{Experimental estimation of the deformation}
\emph{Noise characterization.} Before performing the experimental 
estimation of small deformations, we need to verify the veracity of 
the assumption of Gaussian noise made in this work for our setup. 
For this, we performed a set of characterization
measurements for two values of the input power and the four fields used 
to perform the estimation. 
The first one was performed at the highest input power possible 
without saturating any pixels of the camera and taking care to 
send the same amount of energy for each 
input field. 
The second one was performed using only 20\% of the total power 
used for the first one. For each set, we sent 500 copies of the 
four fields and used the measured  output intensity distributions to compute the 
mean and standard deviation value of the intensity
over the output pixels as shown 
in \refedfig{fig:noise}. At higher power we can clearly see some 
intensity-dependent behavior in the standard deviation which 
is used to characterize the noise, particularly for the OCC for 
which the focal spot is clearly visible. At lower power, however, 
this dependence on the intensity for the noise is almost completely gone
with the $\sigma_n$ values being fairly uniform. These results 
confirm that our assumption of Gaussian noise are valid particularly 
at the powers used to perform the estimation. 


\emph{Estimating the deformation.}
Given that the perturbations under consideration are small, it is sensible
to assume a linear model for the measured intensity distribution $\bs \chi$
over the output modes after a change $\Delta \xi$ in the perturbation.
Explicitly, we have that
\begin{align}
  \bs{\chi}  \approx  \bt I (\zeta_i) + \bt I_\zeta (\zeta_i)
  \Delta \zeta + \bs{\mathscr{w}}(\sigma),
\end{align}
where $\bt I(\zeta_i)$ represents the output intensity distribution over
the output modes prior to changing the deformation, 
$\bt I_\zeta (\zeta_i)$ is the derivative of $\bt I$ 
with respect to $\zeta$ and evaluated at $\zeta_i$, and
$\bs{\mathscr{w}}(\sigma)$ is a vector of normally distributed independent 
random variables with zero mean and standard deviation $\sigma$. 
For this linear model the minimum variance unbiased estimator 
\cite{sengupta1995fundamentals} is given by
\begin{align}\label{eq:est}
  \Delta \zeta^{(\text{est})}(\bs{\chi}) = 
  \frac{\bt I_\zeta (\zeta_i) \cdot 
  \left[ \bs{\chi}  - \bt I (\zeta_i)\right]}
  {\norm{\bt I_\zeta (\zeta_i) }^2}.
\end{align}

In this model, both the reference intensity distribution $\bt I$ 
and the derivative $\partial_\zeta \bt I$ are assumed to be 
known. 
To calibrate the derivative, we first set the input power as high as possible 
without saturating any pixels for any of the fields used for the estimation. 
Then, we perform 500 independent measurements for each field and for two
deformations centered around the reference value $\zeta_i$. Then, the derivative 
of the intensity distribution of each field is estimated by taking the difference 
between the mean for each deformation and dividing by the total deformation performed
between the two points. As can be seen in \refedfig{fig:chgI}, the distribution of
the mean change in intensity obtained from these calibration measurements is 
quite smooth, even though the effect of noise is noticeable for each individual realization, 
and with its main values above
the noise level shown in \refedfig{fig:noise} for all the fields. 
Also shown in \refedfig{fig:chgI} are the distribution of
the mean change in intensity and a single realization at the input power used for the
estimation of the smallest change. Here, we can see that each individual change is quite
susceptible to noise and that its effect is even noticeable 
for the mean distribution, particularly
for the worst performing field given by the random wavefront. 
For the reference intensity distribution, we can use the same 
input power used for the estimations
since the mean intensity distribution is well above the noise 
level for all fields (see \refedfig{fig:noise}). Moreover, 
using the same input power for the reference as for the 
estimation allows the nonuniform background to be automatically removed.

\section{Data availability}

Examples of the data can be found in the dedicated GitHub repository \citesupp{gutierrez2023repo}.
All other datasets used in this work are available from the authors upon
reasonable request.

\section{Code availability}

All the code used to produce the results and 
figures presented in this work can be found
in the dedicated GitHub repository~\citesupp{gutierrez2023repo}.

\bibliographystylesupp{arthur}
\bibliographysupp{refs}

\section{Acknowledgements}
R.~G.~C. acknowledges  L.~L.~S\'anchez-Soto and  K.~Liang for useful discussions. 

\section{Author contributions}
R.G.C. conceived the project, developed the theory and performed 
the experimental measurements. S.M.P. supervised the experimental
work and theory development. 
All authors contributed to the scientific discussion and the writing of the manuscript. 

\section{Funding} 
The authors acknowledge the French Agence Nationale pour la Recherche 
(Grant No. ANR-20-CE24-0016 MUPHTA) and the Labex WIFI 
(ANR-10-LABX-24, ANR-10-IDEX-0001-02 PSL*).

\section{Competing interests}
The authors declare no competing interests.

\balance


\setcounter{figure}{0}   
\renewcommand{\figurename}{Extended Data Figure}
\begin{figure*}
  \includegraphics[width=.55\linewidth]{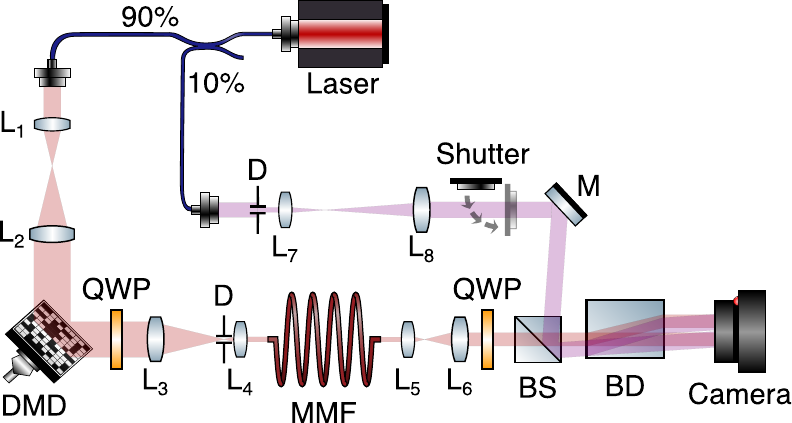}
  \centering
  \figurecaption{Optical setup}{DMD, digital micromirror device;
  L\textsubscript{i}, lenses; QWP, quarter-wave plates; M, mirror; D, diaphragm; 
  BD, beam displacer; BS, beam splitter. The shutter is used to block the 
  reference beam in order to perform intensity measurements.}
  \label{fig:setup} 
\end{figure*}

\begin{figure*}
  \includegraphics[width=.55\linewidth]{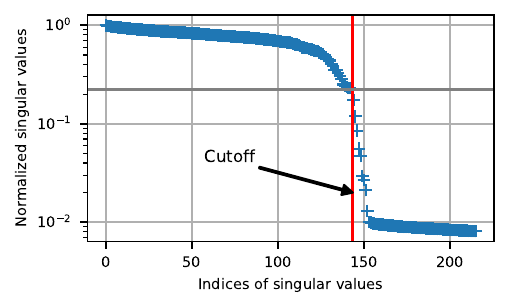}
  \centering
  \figurecaption{Identifying the fiber modes}{Singular values of the TM measured in the pixel basis. The fiber modes are identified as the singular vectors for which the singular values light in the plateau. The cutoff value is marked by the vertical red line.}
  \label{fig:svdpix} 
\end{figure*}

\begin{figure*}
  \includegraphics[width=.99\linewidth]{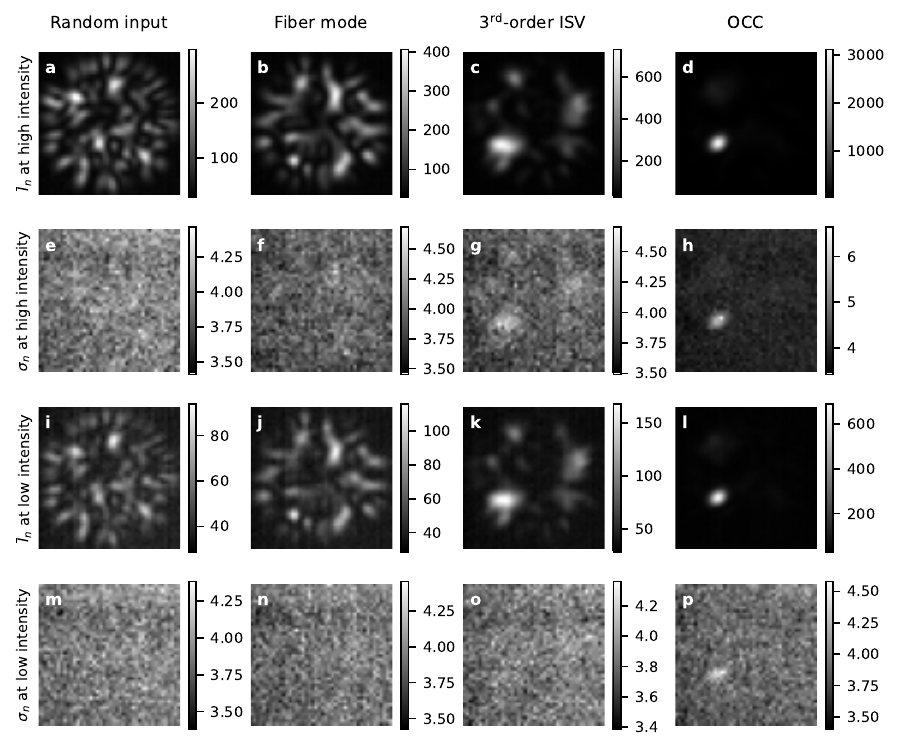}
  \centering
  \figurecaption{Noise characterization}{\textbf{a-d,i-l,} Mean intensity distributions and \textbf{e-h,m-p,} standard deviations for the four fields used for the parameter estimation at high-intensity values (\textbf{a-h}) and when only 20\% of the light is sent (\textbf{i-p}).
  }
  \label{fig:noise} 
\end{figure*}

\begin{figure*}
  \includegraphics[width=.99\linewidth]{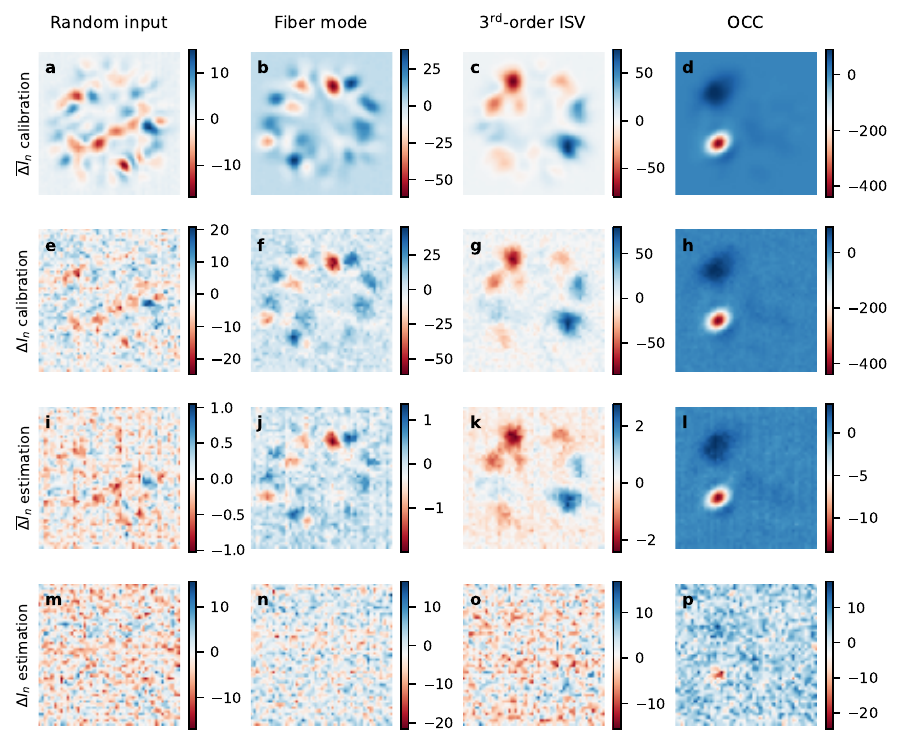}
  \centering
  \figurecaption{Change in intensity}{\textbf{a-d,i-l,} Mean and \textbf{e-h,m-p,} single  change in intensity distributions for the four fields used to calibrate the derivative and acquired at high-intensity values (\textbf{a-h}), and those used to estimate the smallest deformation when using only 20\% of the light sent for the calibration (\textbf{i-p}). 
  }
  \label{fig:chgI} 
\end{figure*}


\hfill 


\end{document}